\documentclass[twocolumn,showpacs,floats,floatfix,superscriptaddress,aps,pra]{revtex4}

\usepackage{amsfonts}
\usepackage{amssymb}
\usepackage{amsmath}
\usepackage{calc}
\usepackage[dvips]{graphicx}
\usepackage{bm}

\def\be{\begin{equation}}
\def\ee{\end{equation}}
\def\bea{\begin{eqnarray}}
\def\eea{\end{eqnarray}}
\def\bse{\begin{subequations}}
\def\ese{\end{subequations}}

\def\1{\mathbf{1}}

\begin{document}

\title{Quantum gate in the decoherence-free subspace of trapped ion qubits} \pacs{03.67.Lx, 37.10.Ty, 32.80.Qk}

\author{}
\author{Peter A. Ivanov}
\affiliation{Institut f\"ur Physik, Johannes
Gutenberg-Universit\"at Mainz, 55099 Mainz, Germany}
\affiliation{Department of Physics, Sofia University, James
Bourchier 5 blvd, 1164 Sofia, Bulgaria}
\author{Ulrich G. Poschinger}
\affiliation{Institut f\"ur Physik, Johannes
Gutenberg-Universit\"at Mainz, 55099 Mainz, Germany}
\author{Kilian Singer}
\affiliation{Institut f\"ur Physik, Johannes
Gutenberg-Universit\"at Mainz, 55099 Mainz, Germany}
\author{Ferdinand Schmidt-Kaler}
\affiliation{Institut f\"ur Physik, Johannes
Gutenberg-Universit\"at Mainz, 55099 Mainz, Germany}

\begin{abstract}
We propose a geometric phase gate in a decoherence-free subspace
with trapped ions. The quantum information is encoded in the
Zeeman sublevels of the ground state and two physical qubits to
make up one logical qubit with ultra long coherence time. Single-
and two-qubit operations together with the transport and splitting
of linear ion crystals allow for a robust and decoherence-free
scalable quantum processor. For the ease of the phase gate
realization we employ one Raman laser field on four ions
simultaneously, i.e. no tight focus for addressing. The
decoherence-free subspace is left neither during gate operations
nor during the transport of quantum information.
\end{abstract}

\maketitle

\section{Introduction}

Trapped ions are among the most promising physical systems for
implementing quantum information due to their long coherence time
as compared with the times required for quantum logic operations
\cite{BW}. A robust quantum memory is a crucial part of the
realization of an ion trap based quantum computer \cite{NC}. One
may distinguish different possibilities for encoding a qubit in a
trapped ion, either one uses a long lived metastable state and
drives coherent transitions on the corresponding optical
transition \cite{FSKaler}, which sets challenging requirements on
the laser source and ultimately limits the coherence time to the
lifetime of the metastable state. Alternatively, a qubit can be
encoded in sublevels of the electronic ground state. This may be
either hyperfine ground state levels \cite{Wineland} or Zeeman
ground states \cite{U} which are coherently manipulated by means
of stimulated Raman transitions. For conventional single-ion
qubits encoded in the Zeeman sublevels as in $^{40}$Ca$^{+}$ are
less favorable, as compared with hyperfine qubits in magnetically
insensitive clock states \cite{Benhelm,CLanger}, as their energy
splitting depends linearly on the magnetic field. Already small
magnetic field fluctuations of about 0.2 T limit the coherence to
250 $\mu$s \cite{JHome}.

We follow in our proposal an elegant alternative
\cite{Aolita,CRoos1} to boost the robustness of such qubits by
using a decoherence-free subspace (DFS)
\cite{KMW,DFS,Z,Hafner,CRoos}. We employ odd Bell states as the
computational basis of \textit{logical qubits} $|0\rangle \equiv
\left\vert \uparrow_1 \downarrow_2 \right\rangle $ and $|1\rangle
\equiv \left\vert \downarrow_1 \uparrow_2 \right\rangle $ with the
overhead of having two physical spin qubits. Ground states
$|{\uparrow}\rangle$ and $|{\downarrow}\rangle$ do not perform any
bit flip errors. Magnetic field or laser phase fluctuation would
lead to errors for a single ion, but in the chosen Bell states
such fluctuations are identical for both ions in the logical
qubit. This assures that such states can maintain coherent of up
to 20~s and single qubit operations in DFS, have been demonstrated
\cite{F,W}. A universal set of single and two qubit operations
between logical qubits has been proposed \cite{Hafner} and
recently performed with a fidelity of $89\%$ \cite{Monz}, and it
would be desirable to reach a fidelity of better than 99$\%$ also
for DFS gates.
\begin{figure}[tb]
\includegraphics[angle=0,width=65mm]{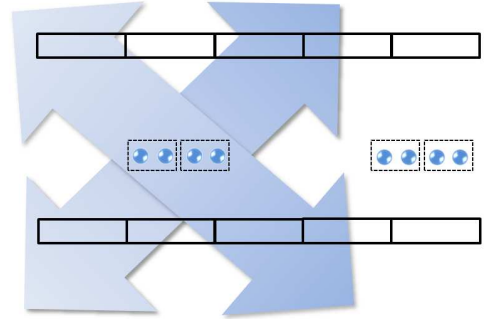}
\caption{(color online). Sketch of a segmented ion trapping device
holding logic qubits in the DFS (dashed boxes). The qubits are
moved to specific zones, merged or separated by applying
time-dependent trap control voltages. A pair of logical qubits
collectively interact with the laser fields for implementing a two
qubit gate. Quantum coherence in the DFS is preserved during the
gate operation and the transports.}\label{fig1}
\end{figure}

In this paper we show how these two-qubit gate operations could be
improved by a novel scheme, which do not require individual ion
addressing. The single ion addressing was identified as one of the
major difficulties and one of the main source of loss in fidelity
in the experiment. Our scheme is based on \emph{homogeneous}
illumination of the four ions. Additionally, our proposed gate
operates in a Raman type of laser excitation between ground state
DFS logic qubits, an additional promising a fidelity improvement
as compared with the metastable, thus 1.2~s long lived basis
states in Ref.~\cite{Monz}. We carefully investigated all
contributions to the spin-dependent light force
\cite{Leibfried,Sackett,Kim,BKRB,Steane}, and optimized the
scheme. Furthermore, we generalize the scheme for a scalable
approach of quantum computing, Fig.\ref{fig1}. With gate times of
about 25 $\mu$s, more that 5 orders of magnitude faster the
decoherence time, the required overhead to handle two ions for one
qubit appears to be relatively small. Even more important, such
favorable separation of time scales would pave the way to realize
quantum error correction \cite{JC}. It further allows for
transporting ion crystals in segmented micro ion traps
\cite{Schulz,CL,JMA} for the creation of cluster states \cite{RB}
and operations with a large number of ions for ultimate
scalability of quantum information processing.

The present paper is organized as follows: Sec. \ref{2} describes
the optimized scheme of the DFS gate between two logical qubits.
In Sec. \ref{3} we show that the gate is suitable for the scalable
gate operations of many logical qubits and specifically for the
creation of cluster states without leaving the DFS. In Sec.
\ref{4} we analyze several error sources relevant to an
experimental implementation of our method. Some of the errors stem
from a single gate operation, some of them occur when transporting
logical qubits to achieve scalability. Finally, in Sec. \ref{5} we
give a summary of the results.

\section{State-dependent force}\label{2}

\begin{figure}[tb]
\includegraphics[angle=0,width=90mm]{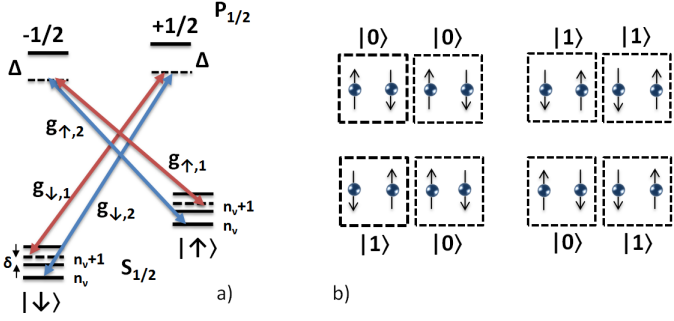}
\caption{(color online) a) Level scheme of a $^{40}$Ca$^{+}$. The
qubit is encoded in the Zeeman sublevels $\left\vert
m_{J}=-1/2\right\rangle =\left\vert \downarrow \right\rangle $ and $%
\left\vert m_{J}=1/2\right\rangle =\left\vert \uparrow
\right\rangle $ of the S$_{1/2}$ ground state. S$_{1/2}$ $\rightarrow$
P$_{1/2}$ transition is driven by a pair of laser beams. Each of
the lasers is circularly polarized with balanced $\sigma^{+}$ and
$\sigma^{-}$ components. $ g_{s_{i},a}$ is the single beam Rabi
frequency of the coupling between the ground states
$s_{i}=\downarrow, \uparrow$ and the excited state of the $i$th
ion. b) Logical qubits basis states are indicated by the dashed boxes.
When the spins of middle two ions are aligned in opposite
directions the light force driving the gate cancels. If and only if the middle two spins
are aligned in the same directions the force pushes the ions in
different direction.} \label{fig2}
\end{figure}

We consider a linear crystal of four ions confined in a linear
Paul trap with trap frequency $\omega_{z}$. The qubit is encoded
in the Zeeman ground states levels $\left\vert \uparrow
\right\rangle =$ $\left\vert m_{J}=1/2\right\rangle $ and
$\left\vert
\downarrow \right\rangle =\left\vert m_{J}=-1/2\right\rangle $ of the S$%
_{1/2}$ ground state of the ion \cite{U}. The linear ion crystal
simultaneously interacts with two non-copropagating laser beams
with frequency difference $\omega _{p}+\delta $, where $\omega
_{p}$ is the $p$th vibrational frequency of the ion crystal and
$\delta $ is the detuning from the vibrational frequency ($\omega
_{p}\gg \left\vert \delta \right\vert $). In contrast to the
center-of-mass mode, the higher energy vibrational modes are less
sensitive to the heating due to fluctuating ambient electric field
because it requires short-wavelength components of the field to
heat it \cite{Turchette}. The laser is detuned
from the S$_{1/2}$ $\rightarrow $P$_{1/2}$ transition with large detuning $%
\Delta $ and couples only the vibrational levels for each of the
spin states according to Fig. \ref{fig2}a.
The interaction Hamiltonian for a string of four ions
simultaneously interacting with a single laser pulse in the
Lamb-Dicke limit and rotating wave approximation is given by
\cite{Monroe}
\begin{equation}
\hat{H}_{I}\left( t\right) =\sum_{\mathbf{s}_{i}}\left( F_{\mathbf{s}%
_{i}}^{(p)}z_{p}e^{-\text{i}\delta t}\hat{a}^{\dag }+F_{\mathbf{s}%
_{i}}^{(p)\ast }z_{p}e^{\text{i}\delta t}\hat{a}\right) \left\vert \mathbf{s}%
_{i}\right\rangle \left\langle \mathbf{s}_{i}\right\vert .
\label{Hamiltonian}
\end{equation}%
Here $\hat{a}$ and $\hat{a}^{\dag }$ are the creation and
annihilation operators of phonons in the $p$th vibrational mode,
$z_{p}=\sqrt{\hbar /2M\omega _{p}}$ is the spread of the ground
state wavepacket for the respective vibrational mode, $M$ is the
ion mass and $\mathbf{s}_{i}=\left\{
s_{1},s_{2},s_{3},s_{4}\right\} $ runs over all spin
configurations of the four ions. The absolute static ac Stark
shift of the energies of the qubit states is given by
$\chi_{s_{i}}=-(\left\vert g_{s_{i},1}\right\vert^{2}+\left\vert
g_{s_{i},2}\right\vert^{2})/2\Delta$, where $%
g_{s_{i},a}$ ($a=1,2$) is the Rabi frequency pertaining to single
beam, Fig. \ref{fig2}a. This shift is generally different for the
qubit states $|{\uparrow}\rangle $ and $|{\downarrow}\rangle $,
thereby induces additional phase in the qubit evolution. The
spatiotemporally varying differential shift, which gives rise to
the spin-dependent force is $\Omega _{s_{i}}=g_{s_{i},1}^{\ast
}g_{s_{i},2}/\Delta $. For the E-mode ($\omega
_{3}\approx\sqrt{5.81}\omega _{z}$) the first and fourth ions
oscillate out of phase and with equal amplitudes with the second
and third ions \cite{James}. Therefore, the magnitude of the
laser-ion coupling is the same for all four ions, but opposite in
sign with respect to the middle two ions. The force on the
collective spin states due to differential Stark shift for the
E-mode is given by
\begin{equation}
F_{s_{1},s_{2},s_{3},s_{4}}^{(3)}=\frac{\hbar \Delta k}{2}\left(
\Omega _{s_{1}}e^{\text{i}\zeta _{1}}-\Omega
_{s_{2}}e^{\text{i}\zeta _{2}}-\Omega _{s_{3}}e^{\text{i}\zeta
_{3}}+\Omega _{s_{4}}e^{\text{i}\zeta _{4}}\right) , \label{Force}
\end{equation}%
where $\Delta k$ is the laser wave vector difference along the
trap axis. The position-dependent
phase is equal to $\zeta _{i}=\Delta kz_{i}^{0}-\Delta \phi $, where $%
z_{i}^{0}=lu_{i}$ with $u_{i}$ is the dimensionless equilibrium
position of the $i$th ion and $l^{3}=Z^{2}e^{2}/4\pi \epsilon
_{0}M\omega _{z}^{2}$ is the length scale parameter. $\Delta \phi
$ is the phase difference between the driving fields. The unitary
operator for the Hamiltonian (\ref{Hamiltonian})
is given by%
\begin{equation}
\mathbf{\hat{U}}_{0}\left( t\right) \mathbf{=}\prod\limits_{\mathbf{s}_{i}}%
\hat{D}\left( \alpha _{\mathbf{s}_{i}}\right) e^{\text{i}\Phi _{\mathbf{s}%
_{i}}},  \label{Propagator}
\end{equation}%
where
\begin{equation}
\hat{D}\left( \alpha _{\mathbf{s}_{i}}\right) =\exp \left[ \left( \alpha _{%
\mathbf{s}_{i}}\hat{a}^{\dag }-\alpha _{\mathbf{s}_{i}}^{\ast }\hat{a}%
\right) \right] \left\vert \mathbf{s}_{i}\right\rangle \left\langle \mathbf{s%
}_{i}\right\vert   \label{D}
\end{equation}%
is the state-dependent displacement operator with
\begin{equation}
\alpha _{\mathbf{s}_{i}}=\left( e^{-\text{i}\delta t}-1\right) F_{\mathbf{s}%
_{i}}^{(3)}z_{3}/\hbar \delta.   \label{displacement}
\end{equation}%
The state-dependent geometric phase
\begin{equation}
\Phi _{\mathbf{s}_{i}}=\left( \delta t-\sin \delta t\right) \left\vert F_{%
\mathbf{s}_{i}}^{(3)}z_{3}\right\vert ^{2}/\left\vert \hbar \delta
\right\vert ^{2}  \label{GPHASE}
\end{equation}%
appears due to non-commutativity of the interaction Hamiltonian at
different times. We project the unitary operator
(\ref{Propagator}) onto the DFS under consideration: \{$\left\vert
\uparrow \downarrow \uparrow \downarrow \right\rangle $,
$\left\vert \downarrow \uparrow \uparrow \downarrow \right\rangle
$, $\left\vert \uparrow \downarrow \downarrow \uparrow
\right\rangle $, $\left\vert \downarrow \uparrow \downarrow
\uparrow \right\rangle $\}. These states are immune to collective
dephasing caused by magnetic field fluctuations.
\begin{figure}[tb]
\includegraphics[angle=-90,width=70mm]{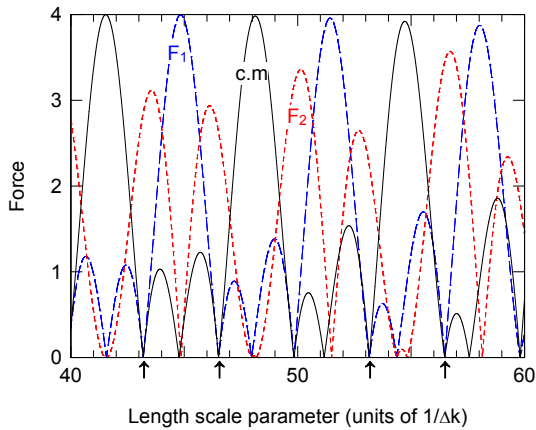}
\caption{(color online). Absolute values of the spin-dependent
forces $ F_{1}=F_{\uparrow \downarrow \uparrow \downarrow
}^{(3)}=F_{\downarrow \uparrow \downarrow \uparrow }^{(3)}$
(dashed) and $F_{2}=F_{\downarrow \uparrow \uparrow \downarrow
}^{(3)}=F_{\uparrow \downarrow \downarrow \uparrow }^{(3)}$
(dotted), Eq. (\ref{Force}) as a function of the length scale
parameter $\l$ for a gate mediated by the E-mode. Here we assume
that $\Omega _{\uparrow }=-\Omega _{\downarrow }$. The forces
$F_{1}$ vanish for $\Delta kl=2\protect\pi n/(u_{4}-u_{2})$, with
$n$ integer. At the same points the forces $F_{\uparrow \downarrow
\downarrow \uparrow }^{\left( 1\right) }$ and $F_{\downarrow
\uparrow \uparrow \downarrow }^{\left( 1\right) }$ (solid) for the
center of mass mode (in brief c.m) are zero, as indicated by the
arrows.} \label{fig3}
\end{figure}
We adjust the trap potential such that
\begin{eqnarray}
\Delta k\left( z_{4}^{0}-z_{1}^{0}\right)  &=&\Delta k\left(
z_{2}^{0}-z_{1}^{0}\right) +2n\pi ,  \notag \\
\Delta k\left( z_{4}^{0}-z_{2}^{0}\right)  &=&\Delta k\left(
z_{3}^{0}-z_{1}^{0}\right) =2n\pi ,  \label{condition}
\end{eqnarray}%
where $n$ is integer. This optimizes the motional coupling to the
E-mode as can be seen from Eq. (\ref{Force}). In Fig. \ref{fig3}
we plot the spin-dependent forces for the E-mode as a function of
the length scale parameter $l$. The spin-dependent forces
$F_{\uparrow \downarrow \uparrow \downarrow }^{(3)}$ and
$F_{\downarrow \uparrow \downarrow \uparrow }^{(3)}$ vary as a
function of $\Delta kl$ and vanish for $\Delta kl=2\pi
n/(u_{4}-u_{2})$, while the
forces $F_{\uparrow \downarrow \downarrow \uparrow }^{(3)}$ and $%
F_{\downarrow \uparrow \uparrow \downarrow }^{(3)}$ displace the
motional state in opposite directions. At the same point two of
the spin-dependent forces for the center-of-mass mode are zero,
hence the off-resonant excitations of this mode do not take place.
During the time evolution the motional state moves along a
circular path in phase space and returns to the origin after time
$T_{g}=2\pi /\delta $, while the spin states acquire geometric
phases $\Phi _{\mathbf{s}_{i}}=2\pi \left\vert F_{\mathbf{s}%
_{i}}^{(3)}z_{3}\right\vert ^{2}/\left\vert \hbar \delta
\right\vert ^{2}$. In this case the force displaces the ions if
the spins of the middle two ions are aligned in the same direction
and cancel if the spins are opposite, Fig. \ref{fig2}b. After time
$T_{g}$ the states for which the force is not canceled acquire
geometric phases $\Phi _{\downarrow \uparrow \uparrow \downarrow
}=\Phi _{\uparrow \downarrow \downarrow \uparrow }$. By proper
choice of the Rabi frequencies $\Omega _{\uparrow }$ and $\Omega
_{\downarrow }$ one can adjust the geometric phases to be $\pi /2$
and the action of the gate onto the DFS states is given by
\begin{eqnarray}
\left\vert \uparrow \downarrow \uparrow \downarrow \right\rangle
&\rightarrow &\left\vert \uparrow \downarrow \uparrow \downarrow
\right\rangle ,  \notag \\
\left\vert \downarrow \uparrow \uparrow \downarrow \right\rangle
&\rightarrow &\text{i}\left\vert \downarrow \uparrow \uparrow
\downarrow
\right\rangle ,  \notag \\
\left\vert \uparrow \downarrow \downarrow \uparrow \right\rangle
&\rightarrow &\text{i}\left\vert \uparrow \downarrow \downarrow
\uparrow
\right\rangle ,  \notag \\
\left\vert \downarrow \uparrow \downarrow \uparrow \right\rangle
&\rightarrow &\left\vert \downarrow \uparrow \downarrow \uparrow
\right\rangle .  \label{Gate}
\end{eqnarray}%
The DFS gate (\ref{Gate}) is a controlled-phase gate between two
logical decoherence-free qubits. Hence, the unitary evolution
transforms any superposition of the states belonging to the DFS
into another superposition of those states. Note that the gate
operation does require an ion localization well within the Lamb
Dicke regime but no ground state cooling of the gate mode, similar
to the geometric phase two-qubit gate \cite{Leibfried}.

Additionally to the geometric phase the qubit states acquire extra
phase due to ac Stark shift $\chi_{s_{i}}$ so that we have
$|{\downarrow}\rangle\rightarrow
e^{-\text{i}\varphi_{\downarrow}}|{\downarrow}\rangle$ and
$|{\uparrow}\rangle\rightarrow
e^{-\text{i}\varphi_{\uparrow}}|{\uparrow}\rangle$, where
$\varphi_{s_{i}}=\int_{0}^{t}\chi_{s_{i}}dt$. The phase is
proportional to the laser intensity thereby it is making the gate
implementation very sensitive to laser-intensity fluctuations.
However, as long as the four ion chain is addressed uniformly and
the quantum information is encoded in the DFS the ac Stark shift
causes only global phase in Eq. (\ref{Gate}). The effect of
slightly inhomogeneous illumination of the ions - accounting for
the experimental reality - and the implication of the gate
fidelity will be discussed in Section \ref{4}
\begin{figure}[tb]
\includegraphics[angle=-90,width=70mm]{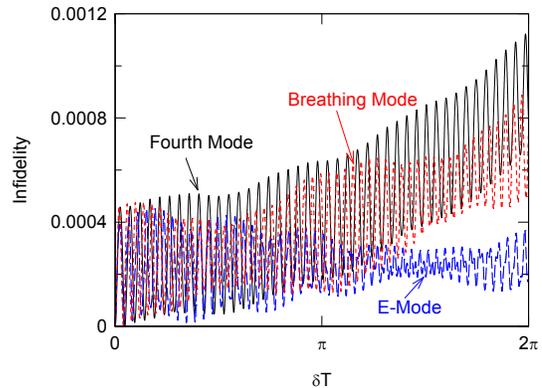}
\caption{(color online). The part of the infidelity $1-F$ due to
off-resonant excitations of the parasitic vibrational modes as a
function of $\protect\delta t$. The string of four ions is
simultaneously addressed with laser fields with frequency close to
the breathing mode (dotted), E-mode (dashed), and fourth mode
(solid). The Rabi frequencies and the axial trap frequencies are
listed in Table \ref{Table}. } \label{fig4}
\end{figure}
\section{Scalability and creation of a linear cluster state}\label{3}
In general, two approaches for scalable quantum computing with ion
string are viable. The one aims for a long ion crystal, where all
ions share common modes of vibration which allow to drive gate
operations between them. The seminal paper by Cirac and Zoller
\cite{JICirac} proposed axial modes and more recently it has been
proposed to use the radial modes of a large ion crystal
\cite{SZhu}. The fundamentally different approach aims to shuttle
ions in segmented traps \cite{KMW,Leibrandt}, such that only two
logical qubits are in the processor region during the quantum gate
operation.

If the second approach is joined with gate operations in DFS and
logical qubits, a maximum of four ions is kept in the central
processor unit and pairs of ions are shuttled together using the
control segments of the trap (\ref{fig1}). A large separation of
time scales for a gate with about 100 ms transport operations are
viable in the ms range \cite{Huber}, as well as a obtaining error
syndromes if one aims for quantum error correction, where the
readout of fluorescence is necessary. In order to implement a
multi qubit controlled phase gate we can decompose it in a
sequence of single qubit gates and geometric phase gates between
the logical qubits \cite{Barenco}. The single logical qubits (pair
of ions) can be trapped in the storage zone and then can be
transported to the processing zone where the gate operations are
performed, Fig. \ref{fig1}.
For instance, a three bit decoherence-free geometric gate is
decomposed as $G_{3}=T_{2,0}G_{3,0}T_{2,0}$. Here $T_{2,0}$ is the
Toffoli gate between the first two control logical qubits and the
auxillary logical qubit which is prepared initially in the state
$|0\rangle$. The second operation $G_{3,0}$ is the geometric phase
gate between the target logical qubit and the auxillary qubit. The
final Toffoli gate is applied to restore the auxillary qubit in
the state $|0\rangle$. The Toffoli gate also can be decomposed as
a succession of Hadamard gates and geometric phase gates between
the logical qubits. Therefore any of the gate operations is
realized with two ions (single qubit operations) and with four
ions (two qubit operations) where the proposed method can be
applied. The generalization to n-bit geometric phase gate is
straightforward: $G_{n}=T_{n-1,0}G_{n,0}T_{n-1,0}$.

As an example we describe the method for the generation of a
linear cluster state. Cluster states are highly entangled states,
which are the fundamental resource of the one-way quantum computer
\cite{RB}. Cluster states have been experimentally created with
atoms in optical lattice \cite{Mandel} and with photons
\cite{Zeilinger}. Multi-qubit cluster states have yet not been
created with trapped ions. An ion-trap architecture for high-speed
measurement-based quantum computer was proposed \cite{SJ}. Ref.
\cite{Ivanov} proposed an efficient technique for the creation of
four, five, and six qubit linear cluster states by collective
bichromatic interaction, while in \cite{Wunderlich} creation of
two-dimensional cluster state was suggested by using a spin-spin
coupling induced by a magnetic-field gradient. Here, we propose
the creation of four linear cluster state, without leaving the DFS. If the gate (\ref%
{Gate}) is applied onto the decoherence-free initial state
$\left\vert \Psi _{in}\right\rangle
=|B_{1,2}\rangle|B_{3,4}\rangle$, which is a product of two Bell
states
$|B_{i,j}\rangle=(|{\uparrow_{i}\downarrow_{j}}\rangle+|{\downarrow_{i}\uparrow_{j}}\rangle)/\sqrt{2}$,
then the geometric phase gate transforms the state to
\begin{equation}
\left\vert \Psi _{c}\right\rangle =(\left\vert \uparrow \downarrow
\uparrow \downarrow \right\rangle +\text{i}\left\vert \uparrow
\downarrow \downarrow \uparrow \right\rangle +\text{i}\left\vert
\downarrow \uparrow \uparrow \downarrow \right\rangle +\left\vert
\downarrow \uparrow \downarrow \uparrow \right\rangle )/2.
\label{cluster}
\end{equation}%
The final state is highly entangled four qubit cluster state.
Since the spin states are not mixed due to the unitary evolution
(\ref{Propagator}), the initial state is transformed into the
final cluster state without leaving the DFS. Combining the ion
trap architecture and the transport of such states in a segmented
trap, (see Fig. \ref{fig1}) large cluster states can be created by
fusing several four linear cluster states with the DFS gate.

\section{Optimizing the gate robustness and remaining errors}\label{4}

During the gate evolution on the axial E-mode, also the other
vibrational modes are off-resonantly excited there might be a
non-vanishing coupling strength; however all radial motional
degrees of freedom of the four ion crystal are excluded by the
geometry of the laser excitation. We try to optimize the gate
fidelity by analyzing and avoiding spurious ac light force driving
on all other modes. For general coupling strengths and $\delta
_{p}=\omega _{p}-\omega _{3}-\delta $, the displacement and
geometric phase associated with the $p$th vibrational mode is
given by
\begin{eqnarray}
\alpha _{\mathbf{s}_{i}}^{(p)}(t) &=&\frac{F_{\mathbf{s}_{i}}^{(p)}z_{p}}{%
\hbar \delta _{p}}\left( e^{-\text{i}\delta _{p}t}-1\right) ,  \notag \\
\Phi _{\mathbf{s}_{i}}^{(p)}(t) &=&\left\vert \frac{F_{\mathbf{s}%
_{i}}^{(p)}z_{p}}{\hbar \delta _{p}}\right\vert ^{2}\left[ \delta
_{p}t-\sin \left( \delta _{p}t\right) \right] .  \label{alfa}
\end{eqnarray}%
\begin{figure}[tb]
\includegraphics[angle=-90,width=70mm]{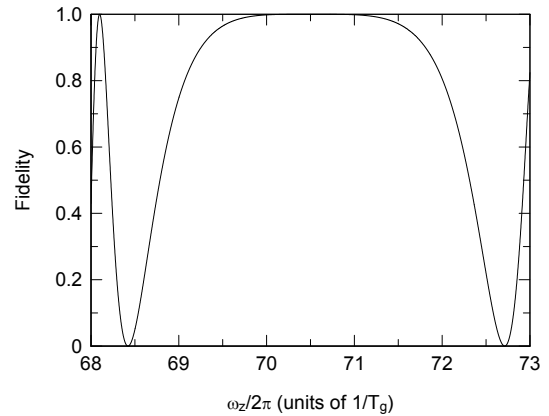}
\caption{The fidelity $F$ Eq. (\ref{Fidelity}) as a function of
the axial trap frequency $\protect\omega _{z}$ for a gate Eq.
(\ref{Gate}) mediated by the E-mode.} \label{fig5}
\end{figure}
As now the absolute force magnitude and the final time are fixed
such that the geometric phase gate condition is fulfilled,
$T_{g}=2\pi /\delta _{3}$, one obtains parasitic displacement and
geometric phases
\begin{eqnarray}
\alpha _{\mathbf{s}_{i}}^{(p)}(T) &=&\frac{F_{\mathbf{s}_{i}}^{(p)}z_{p}}{%
\hbar \delta _{p}}\left( e^{-\text{i}2\pi \delta _{p}/\delta
_{3}}-1\right) ,
\notag \\
\Phi _{\mathbf{s}_{i}}^{(p)}(T) &=&\left\vert \frac{F_{\mathbf{s}%
_{i}}^{(p)}z_{p}}{\hbar \delta _{p}}\right\vert ^{2}\left[ 2\pi
\delta _{p}/\delta _{3}-\sin \left( 2\pi \delta _{p}/\delta
_{3}\right) \right] . \label{alfa}
\end{eqnarray}%
An appropriate measure of the gate fidelity is given by
\cite{KOSLOFF}:
\begin{equation}
F=\frac{1}{N^{2}}\sum_{i,j}\langle
\mathbf{\tilde{s}}_{i}|\hat{U}_{0}^{\dagger
}\hat{U}|\mathbf{\tilde{s}}_{i}\rangle \langle
\mathbf{\tilde{s}}_{j}|\hat{U}^{\dagger
}\hat{U}_{0}|\mathbf{\tilde{s}}_{j}\rangle , \label{fidelity}
\end{equation}%
where $N=4$ is the dimension of the DFS and $i,j$ each run over
all spin configurations belonging to DFS. $\hat{U}_{0}$ is the
desired unitary transform given
by Eq. \ref{Gate} and $\hat{U}$ is the actual one. Neither $\hat{U}_{0}$ nor $%
\hat{U}$ is mixing the spin states. The action of $\hat{U}$ on any
basis state is simply given by
\begin{equation}
|\mathbf{\tilde{s}}_{i}\rangle=|\mathbf{s}_{i}\rangle |vac_{1,2,3,4}\rangle \rightarrow e^{\text{i}%
\sum_{p}\Phi _{\mathbf{s}_{i}}^{\left( p\right)
}}|\mathbf{s}_{i}\rangle
\left\vert \alpha _{\mathbf{s}_{i}}^{(1)}\ldots \alpha _{\mathbf{s}%
_{i}}^{(4)}\right\rangle .
\end{equation}%
where $|vac_{1,2,3,4}\rangle$ indicates the ground state of all axial modes.
The fidelity can then be evaluated under
consideration of the matrix element $\langle 0|\alpha \rangle
=e^{-|\alpha |^{2}/2}$. Hence, we obtain for the fidelity at time
$t$
\begin{equation}
F=\frac{1}{16}\left\vert \sum_{i}\prod_{p}e^{-\frac{1}{2}|\alpha _{\mathbf{s}%
_{i}}^{(p)}|^{2}+\text{i}\Phi _{\mathbf{s}_{i}}^{(p)}}\right\vert
^{2}. \label{Fidelity}
\end{equation}%
\begin{table}[tbp]
\begin{tabular}{|c|c|c|c|c|}
\hline Mode & Breathing & E-mode & Fourth \\ \hline $\omega
_{z}/2\pi $ (MHz) & 2.86 & 2.82 & 2.68\\ \hline $\Omega /2\pi $
(kHz) & 119.64 & 130.62 & 130.47\\ \hline Infidelity & $8.1\times
10^{-4}$ & $1.8\times 10^{-4}$ & $7.7\times 10^{-4}$ \\ \hline
$\Delta \omega _{z}/2\pi $ (kHz) & 53 & 65 & 92 \\ \hline
\end{tabular}
\caption{ The values of the axial trap frequency $\omega _{z}$ and
the Rabi frequency $\Omega $ required for implementation of the
DFS gate for three different vibrational modes. The infidelity due
to the off-resonant excitation is also presented. The frequency
plateaus, where the gate fidelity is better than $99\%$ are listed
in order to elucidate the sensitivity on this parameter.}
\label{Table}
\end{table}

As an example of qubit we consider the $^{40}$Ca$^{+}$ ion with
qubit states encoded in the Zeeman sublevels of S$_{1/2}$ state.
The two-photon Raman transition is driven by a laser field with a
wave length of $\lambda \approx
397$ nm and a wave vector difference along the trap axis $\Delta k\approx2\sqrt{%
2}\pi /\lambda $. In order to cancel the spin forces $F_{\uparrow
\downarrow \uparrow \downarrow }$ and $F_{\downarrow \uparrow
\downarrow \uparrow }$, Eq. (\ref{condition}) we choose distance
parameter $n=15$. The Rabi frequencies are $\Omega _{\uparrow
}=-\Omega _{\downarrow }$ with gate time $T_{g}=25$ $\mu $s and a
laser detuning $\delta =2\pi \times 40$ kHz. Table I lists the
values of the axial trap frequency $\omega _{z}$ and the Rabi
frequency $\Omega $ needed for the synthesis of the DFS gate for
the three different vibrational modes. We compare the minimum gate
infidelity $1-F$ mediated by different vibrational modes due to
the off-resonant excitations, (see, Fig. \ref{fig4}). Even for one
cycle in phase space $\delta T_{g}=2\pi $ the infidelity for the
E-mode is smallest since the off-resonant excitation of the
center-of mass mode for spin states $|{\uparrow \downarrow
\downarrow \uparrow}\rangle$ and $|{\downarrow \uparrow \uparrow
\downarrow}\rangle$ vanishes. 
\begin{figure}[tb]
\includegraphics[angle=-90,width=70mm]{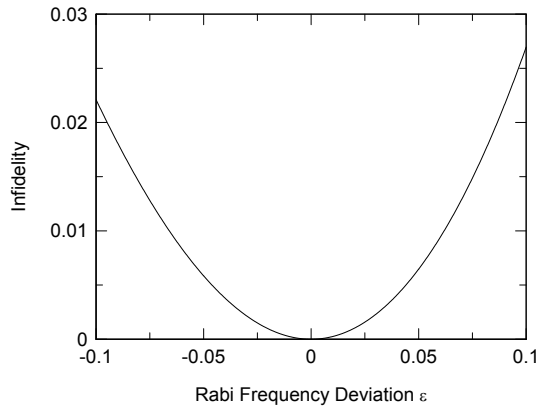}
\caption{Infidelity as a function of the deviation $\protect%
\epsilon $ of the Rabi frequencies defined as $\Omega _{\left\{
s_{i}\right\} }\left( t\right) =$ $\Omega _{\left\{ s_{i}\right\}
}(1+\varepsilon )$ for the implementation of the decoherence-free
gate (\ref{Gate}) mediated by the E-mode.} \label{fig6}
\end{figure}

In Fig.~\ref{fig5} we show the fidelity (\ref{Fidelity}) as a
function of the axial trap frequency $\omega _{z}$ for fixed gate
time $T_{g}$. The Rabi frequency is chosen such that the condition
$\Phi _{\downarrow \uparrow \uparrow \downarrow }=\Phi _{\uparrow
\downarrow \downarrow \uparrow }=\pi/2$ is fulfilled. The
frequency plateaus, where the minimum fidelity is better than
$99\%$ for the gate implementation mediated by the different
vibrational modes are listed in Table \ref{Table}.

Additionally, the proposed gate scheme shows a remarkable
robustness against laser intensity fluctuations. As both laser
beams are derived from the same laser source, we assume that the
intensity fluctuations are common. The corresponding Rabi
frequencies fluctuate therefore simultaneously for all ions. As a
result, we observe that the fidelity only decrease quadratically
with the fluctuations, see Fig.~\ref{fig6}.

The ac Stark shift from the Raman beams can be compensated by a proper choice of polarizations of both beam or by employing an additional compensation beam \cite{FSK2003}. The precision of this compensation is only limited by the spin coherence time, in a typical experiment one might reach an accuracy $\delta_{ac} /(2\pi)$ well below 1~kHz. Such shifts scramble the spin qubit phases as they translate laser intensity fluctuations into effective magnetic field fluctuations. However, the logical qubits are inherently robust against spin phase fluctuations. Errors might only occur when the when an uncompensated ac Stark shift is accompanied by an inhomogeneous illumination of two ions which comprise a logical qubit. As this error of higher order, with a fraction of the above number, we do not take it into further consideration. In a fusing process of three logical qubits for the scalable scheme discussed above, when in the first step two of the qubits are exposed to the gate laser field while the third qubit is not illuminated, even an uncompensated ac Stark shift would not lead to errors as long as the logical qubits are illuminated equally.


Finally, the coherence of qubits in DFS is limited by the
fluctuations of the gradient of the magnetic field over the
distance of the two ion crystal, a few $\mu$m. When transporting
such logical qubits in the trap one may expose them to varying
magnetic gradients, however the phase evolution of the logical
qubits is deterministic \cite{CRoos} and can be easily corrected
for.

The motional decoherence of the vibrational motion of the trapped
ions is the most serious limiting factor in ion trap quantum
information processing. In order to achieve high fidelity the gate
time $T_{g}$ required for the implementation of DFS gate
(\ref{Gate}) must be much shorter than the heating time $\tau $.
The measured heating time of the center of mass mode for
$^{40}$Ca$^{+}$ ion in a segmented micro-ion traps \cite{U} was
found to be 3.3 ms, a typical value for this devices. The heating
time for higher energy modes is expected to be much larger. Hence
the gate time is more than two orders of magnitude faster than the
heating time such that the fidelity of the gate is not affected.

\section{Conclusion}\label{5}

In conclusion we proposed a simple and robust technique for a
decoherence-free controlled phase gate between two logical qubits.
We studied in detail the fidelity of the gate implementation
taking into account various error sources such as off-resonant
transitions, laser fluctuations, and the deviation of the right
choice of the axial trap frequency. We have compared the error
sources for a gate mediated by three different modes, and we have
shown that the gate mediated by the Egyptian mode minimizes the
off-resonant transitions. Our scheme includes the creation of a
linear cluster state within a decoherence free subspace manifold
of four qubits - a starting point for decoherence-free ion trap
one way quantum computing.

\section{Acknowledgments}

This work has been supported by the European Commission projects
EMALI (contract No MRTN-CT-2006-035369), MICROTRAP (contract No.
517675), the Bulgarian NSF grants VU-F-205/06, VU-I-301/07,
D002-90/08, and the Elite programme of the Landesstiftung
Baden-W\"{u}rttemberg.


\end{document}